\documentclass[%
 aip,
 jmp,%
 amsmath,amssymb,
 reprint,%
]{revtex4-1}
\usepackage{graphicx}
\usepackage{dcolumn}
\usepackage{bm}

\usepackage{graphicx}
\usepackage{epsfig}
\usepackage{epstopdf}

\usepackage{listings,xcolor}
\begin{document}


\title[]{Exact low-temperature series expansion for the partition function of the two-dimensional zero-field $s=\frac{1}{2}$ Ising model on the infinite square lattice}

\author{Grzegorz Siudem}
\email{siudem@if.pw.edu.pl}
\author{Agata Fronczak}
\email{agatka@if.pw.edu.pl}
\author{Piotr Fronczak}
\email{fronczak@if.pw.edu.pl}

 \affiliation{Faculty of Physics, Warsaw University of Technology,\\
Koszykowa 75, PL-00-662 Warsaw, Poland}

\date{\today}

\begin{abstract}
In this paper, we provide the exact expression for the coefficients in the low-temperature series expansion of the partition function of the two-dimensional Ising model on the infinite square lattice. This is equivalent to exact determination of the number of spin configurations at a given energy. With these coefficients, we show that the ferromagnetic--to--paramagnetic phase transition in the square lattice Ising model can be explained through equivalence between the model and the perfect gas of energy clusters model, in which the passage through the critical point is related to the complete change in the thermodynamic preferences on the size of clusters. The combinatorial approach reported in this article is very general and can be easily applied to other lattice models.
\end{abstract}

\maketitle

\section{\label{sec:level1}Introduction}

Over the past 100 years, the lattice spin systems were the most actively studied models in statistical mechanics, principally due to their being perhaps the simplest models exhibiting cooperative phenomena, or phase transitions. By far the most important and most extensively studied of these systems is the spin $s=\frac{1}{2}$ Ising model on a square lattice in the absence of an external field, in which each site $i=1,2,\dots V$ has two possible states: $s_i=+1$ or $s_i=-1$. The Hamiltonian of the model can be written in the form
\begin{equation}\nonumber
\mathcal{H}(\{s_i\})=-J\sum_{\langle i,j\rangle}s_is_j,
\end{equation}
where the sum runs over all nearest-neighbour pairs of lattice sites and counts each pair only once, and $-J$ is the energy of a pair of parallel spins. The importance of this model stems from the fact that it belongs to the few models of statistical physics for which exact computations may be carried out (for general reading see \cite{bookBaxter,bookMcCoyWu}).

The first exact, quantitative result for the two dimensional Ising model on a square lattice was obtained in 1941 by Kramers and Wannier \cite{1941Kramers}, who used the low- and high-temperature expansion method to formulate the self-duality transformation by means of which they find the exact critical temperature of the system. Shortly afterwards, in 1944, their result was confirmed by Onsager \cite{1944Onsager}, who derived an explicit expression for the free energy in zero field and thereby established the precise nature of the critical point. And although, at present, the list of different developments in the study of the model is relatively long (for a quick historical overview see preface to the chapter~10 in Ref.~\cite{bookMcCoy}), with this article we complement the list with a new important item: the exact low-temperature series expansion for the partition function of the model on the infinite lattice. To be concrete, we provide the exact expression for the coefficients in the expansion, which is equivalent to exact determination of the number of spin configurations at a given energy. Recently, different issues (both theoretical and computational) related to this problem  have been discussed (see e.g. Refs.~\cite{1991PRLBhanot,1996PRLBeale,2001PRLWang,2007PRLHadbeck,bookLandau,2004PREHaggkvist} and their numerous citations). This discussion has always been more or less clearly associated with an attempt to find an answer to the fundamental question of how signals for phase transitions can be inferred from the number of energy states. In the following, by considering the energy distribution, which is the probability of finding the system in an equilibrium state with a given energy, we shed some light on these issues.

The first lengthy low-temperature series expansion of the partition function per spin for the square lattice Ising model in the absence of the magnetic field was calculated by Domb in 1949 \cite{1949Domb}:
\begin{equation}\label{DombZ}
Z(x)=\frac{2}{x}\left(1\!+\!x^4\!+\!2x^6\!+\!5x^8\!+\!14x^{10}\!+\!44x^{12}\!+\!\dots\right),
\end{equation}
where $x=\exp[-2\beta J]$ and $\beta=(k_BT)^{-1}$. Terms in Eq.~(\ref{DombZ}) were obtained in a systematic way from matrix operators, but the process of their derivation was very tedious and no general expression for the lattice constants (i.e. coefficients in the expansion) was given. In this paper, we use some ideas and formulas, which originate from combinatorics, to get the exact expression for the coefficients. And although our result is important in itself, it is also a pretext to draw physicists' attention to the progress made in recent years in (enumerative) combinatorics \cite{bookComtet,bookStanley}, due to which some theoretical issues related to series expansions in physics of lattice systems \cite{1974bookDomb,2001Orrick,2011Chan} may be treated in a completely different way to provide new insights into the already solved problems and to stimulate yet another actions towards unsolved models.

Although, as far as we know, the Bell-polynomial approach for the Ising model, which is described in this paper, was not considered in the literature, it may be viewed as a variation of the cluster expansion \cite{2010Faris, 2006Kotecky} or Mayer-Ursell formalism  \cite{1927Ursell, 1940Mayer}. The mentioned, well-known techniques provide systematic procedures for the series expansion of the free energy \cite{2006Kotecky}. Coefficients of those series expansions are strictly related to the enumeration of some combinatorial or geometrical structures\cite{2010Faris}. In some sense, our Bell-polynomial approach is an inverse operation to the cluster expansion, because we start with the free energy, which is given as a series, and then calculate coefficients of the series expansions of the partition function.

\section{\label{sec:level2} Derivation of the main result}

The main idea behind this paper is that the low temperature series expansion of the partition function, $Z(x)$, of any lattice model can be easily obtained from the low temperature series expansion of the corresponding free energy, $f(x)$. In this article we consider the Ising model on a square lattice in the so-called bulk version. More specific our calculations based on the Kaufman-Onsager  solution of the model in the case of the periodic boundary conditions. Because of the fact that we analyse  only bulk version of free energy (i.e. free energy per site in the limit of the infinity lattice) our considerations in that point are independent of the chosen boundary conditions.  In the mentioned case the corresponding expression between $Z(x)$ and $f(x)$ can be written in the following form \cite{APF2014}:
\begin{align}\label{apf1}
Z(x)&=2\exp[-\beta f(x)]\\\label{apf2}&=2\exp\left[-\ln x+\sum_{n=1}^\infty a_n\frac{x^n}{n!}\right] \\\label{apf3}&=\frac{2}{x}\left(1+ \sum_{N=1}^{\infty}\frac{1}{N!}Y_N(\{a_n\})x^N\right),
\end{align}
where the factor $2x^{-1}=2e^{2\beta J}$ is due to the doubly degenerate ground state of energy $-2J$, in which all the spins are aligned, and the series coefficients in Eq.~(\ref{apf3}),
\begin{equation}\label{apf4}
g(N)=\frac{1}{N!}Y_N(\{a_n\}),
\end{equation}
which are given by the $N$-th complete Bell polynomials, $Y_N(\{a_n\})$, stand for the number of spin configurations with energy $2JN$ above the ground state. Finally, the complete Bell polynomials in Eqs.~(\ref{apf3}) and (\ref{apf4}) are defined as follows:
\begin{equation}\nonumber
Y_N(\{a_n\})=\sum_{k=1}^N B_{N,k}(\{a_n\}),
\end{equation}
where $B_{N,k}(\{a_n\})$ represent the so-called partial (or incomplete) Bell polynomials, which can be calculated from the expression below:
\begin{equation}\label{bell2}
B_{N,k}(\{a_n\})=N!\sum_{\{c_i\}}\prod_{n=1}^{N-k+1}\frac{1}{c_n!}\left(\frac{a_n}{n!}\right)^{c_n},
\end{equation}
where the summation takes place over all integers $c_n\geq 0$, such that
\begin{equation}\label{bell3}
\sum_{n=1}^{N-k+1}c_n=k\;\;\;\; \mbox{and}\;\;\;\; \sum_{n=1}^{N-k+1}nc_n=N.
\end{equation}

In order to get Eq.~(\ref{apf3}) the generating function for Bell polynomials \cite{bookComtet} has been used, which is equivalent (as far as $a_n\geq 0$ for all $n\geq 0$) to the so-called \emph{exponential formula}, which is a cornerstone of enumerative combinatorics. The formula deals with the question of counting composite structures that are built out of a given set of building blocks \cite{bookWilf}. It states that the exponential generating function for the number of composite structures, $Z(x)$, is the exponential of the exponential generating function for the building blocks, $-\beta f(x)$. Here, it is interesting to note that the famous dimer solutions of the zero-field planar Ising models initiated by Kasteleyn \cite{1961Kasteleyn,1967Kasteleyn}, and further developed by many others (e.g. see papers citing Ref.~\cite{1966Fisher}), are a direct consequence of this formula, in which the partition function stands for the generating function of the number of spin configurations with a given energy, and the free energy is the generating function for dimmers.

Returning to the main topic of this paper: As seen in Eqs.~(\ref{apf1})--(\ref{apf4}), to provide the exact expression for the coefficients $g(N)$ in the low temperature series expansion of the partition function, the coefficients $\{a_n\}$ in the low temperature expansion of $-\beta f(x)$, must first be determined. Starting from the famous result of Onsager for the bulk free energy per site:\\
\begin{widetext}
\begin{equation}\label{f0}
-\beta f(\beta)=-\ln x+\sum_{n=1}^\infty a_n\frac{x^n}{n!}= \ln 2 + \frac{1}{8\pi^2}\int_0^{2\pi}d\theta_1\int_0^{2\pi}d\theta_2 \ln\left[\cosh^2(2\beta J)-\sinh(2\beta J)(\cos\theta_1+\cos\theta_2)\right].
\end{equation}
\end{widetext}
One can show (see Appendix~\ref{sec1a}) that for odd values of $n$ the coefficients are equal to zero:
\begin{equation}\label{An1}
a_n=0,
\end{equation}
while for even values ​​of $n$ they are given by:
\begin{align}\label{An2}
a_n=\frac{1}{2}n!\sum_{d_1,d_2,d_3,d_4}&\binom{d_1+d_2+d_3+d_4}{d_1,\,d_2,\,d_3,\,d_4}\\
\times&\frac{(-1)^{d_2+d_3+d_4-1}2^{d_2}}{d_1+d_2+d_3+d_4}\binom{d_1+d_3}{\frac{d_1+d_3}{2}}^2,\nonumber
\end{align}
where the summation takes place over all quadruple numbers $d_1,d_2,d_3,d_4\geq 0$, which satisfy conditions $d_1+2d_2+3d_3+4d_4=n$ and $d_1+d_3$ is even.

By using Eqs.~(\ref{An1}) and~(\ref{An2}), one gets the following sequence:
\begin{align}
\{a_n\}=&\left\{0,\,0,\,0,\,4!,\,0,\,2\!\cdot\!6!,\,0,\, \frac{9}{2}\!\cdot\!8!,\,0,\right.\\\nonumber&\left.12\!\cdot\!10!,\,0,\,\frac{112}{3}\!\cdot\!12!, \,0,\,130\!\cdot\!14!,\,0,\,\frac{1961}{4}\!\cdot\!16!,\dots\right\},
\end{align}
from which the known expression for the low temperature series expansion of the bulk free energy per site, Eq.~(\ref{f0}), can be drawn (cf. Eq.~(15) in \cite{1996PRLBeale}):
\begin{align}\label{f1}
-\beta f(x)=&-\ln x+x^4+2x^6+\frac{9}{2}x^8+12x^{10}\\\nonumber &+\frac{112}{3}x^{12}+130x^{14}+\frac{1961}{4}x^{16}+\dots.
\end{align}

Up to this point our considerations were exact and concentrated on the bulk case of the infinite square-lattice Ising model. Nonetheless, the presented results may also provide an approximate formulae for the coefficients $g(N,V)$ in the low-temperature series expansion of the partition function for the Ising model on a finite square lattice of the size $V=M\times M$, with the periodic boundary conditions i.e.
\begin{equation}\label{z1}
Z(x,V)\!=\!2\exp[-\beta F(x,V)]\!=\!\frac{2}{x^V}\sum_{N=0}^{V}g(N,V)x^N,
\end{equation}
where $F(x,V)$ stands for the free energy. In this case, we denote series expansion of the free energy as\linebreak $F(x,V)=-V\ln x+\sum_{n=1}^\infty A_n(V)\frac{x^n}{n!}$. 
One can consider the following approximation for the free energy: $F(x,V)\approx Vf(x)$. This approximation provides the exact formula for the coefficients $A_n(V)=Va_n$ with $n<M$, whereas for $M\leqslant n\leqslant V$ the approximation is increasingly less accurate. In the first case, for $n<M$, there are no contours wrapped around the torus. Therefore, the corresponding terms of free energy, $F(x,V)$, are simply proportional to terms in $f(x)$ with $V$ as a proportionality factor. On the other hand, for $n$ comparable and larger than $M$, one must take into account those wrapped contours and the coefficients in the series expansion of $F(x,V)$ and $f(x)$ are no longer proportional. Nevertheless, since the $N$-th Bell polynomial depends only on the first $N$ variables, cf.~Eqs.~(\ref{bell2}) and~(\ref{bell3}), it is true that for~$N<M$:\\
\begin{widetext}
\begin{align}
\label{z2}g(N,V)=&\frac{1}{N!} Y_N\left(0,\,0,\,0,\,V4!,\,0,\,2V\!\cdot\!6!,\,0,\, \frac{9}{2}V\!\cdot\!8!,\,0,\,12V\!\cdot\!10!,\,0,\,\frac{112}{3}V\!\cdot\!12!,\,0,\dots\right),
\end{align}
which allows one to obtain the first terms in the series expansion of Eq. (\ref{z1})
\begin{align}
\label{z3}Z(x,V)&=\frac{2}{x^{V}}\left(1+Vx^4+2Vx^6+\left(\frac{9}{2}V\!+\!\frac{1}{2}V^2\right)x^8+ (12V\!+\!2V^2)x^{10}+\left(\frac{112}{3}V\!+\!\frac{13}{2}V^2\!+\!\frac{1}{6}V^3\right)x^{12}+\dots\right).
\end{align}
\end{widetext}

\section{Discussion}
Now, a few comments about the obtained results are in order. First, we checked numerically that the coefficients in the low temperature series expansion of the free energy are non-negative and grow exponentially as (see Appendix \ref{sec3a})
\begin{equation}\label{Anfig}
\lim_{n\rightarrow\infty}\frac{a_{2n}}{(2n)!}=C\alpha^{2n},
\end{equation}
with $C$ being a positive constant and
\begin{equation}\label{alpha}
\alpha\simeq\frac{1}{x_c}=\exp\left[\frac{2J}{k_BT_c}\right]=\frac{1}{\sqrt{2}-1},
\end{equation}
where $T_c$ is the critical temperature at which the second-order phase transition in the Ising model occurs. The non-negative character of these coefficients is very significant: It brings to mind the so-called \emph{perfect gas of clusters} model \cite{2003Sator}, in which the coefficients, i.e. $\{a_n\}$, stand for the number of microscopic realisations of clusters of size $n$ \cite{APF2014,AF2012,AF2013,GS2013}. For completeness, let us recall that in the perfect gas of clusters model, particles constituting a fluid may interact only when they belong to the same cluster (i.e. there is no potential energy of interaction between the clusters), and the clusters do not compete with each other for volume.

To these ideas have become more intelligible, let us consider $N$ distinguishable elements (particles, portions of energy etc.) partitioned into $k$ non-empty and disjoint subsets (groups, energy clusters etc.) of $n_i>0$ elements each, where $\sum_{i=1}^kn_i=N$. There are exactly
\begin{equation}\label{y1}
{N\choose n_1,\dots,n_k}= N!\prod_{i=1}^k\frac{1}{n_i!}= N!\prod_{n=1}^{N-k+1}\left(\frac{1}{n!}\right)^{c_n}
\end{equation}
of such partitions, where $c_n\geq 0$ stands for the number of subsets of size $n$, with the largest subset size being equal to $N-k+1$, and where Eqs.~(\ref{bell3}) are satisfied. Suppose further that in such a composition, subsets of the same size are indistinguishable from one another, and each of $c_n$ subsets of size $n$ can be in any one of $a_n\geq 0$ internal states. Then the number of partitions becomes:
\begin{equation}\label{y2} N!\prod_{n=1}^{N-k+1}\frac{1}{c_n!}\left(\frac{a_n}{n!}\right)^{c_n}.
\end{equation}
Summing the last expression, Eq.~(\ref{y2}), over all integers $c_n\geq 0$ specified by Eqs.~(\ref{bell3}) one gets the partial Bell polynomial, $B_{N,k}(\{a_n\})$, which is defined by Eq.~(\ref{bell2}). Then, summing the partial polynomials over $k$ one gets the complete polynomial, $Y_N(\{a_n\})$, the combinatorial meaning of which is obvious (i.e. they describe the number of partitions of a set of size $N$ into an arbitrary number of subsets), and whose exponential generating function, $\sum_{N=1}^\infty Y_{N}(\{a_n\})x^N/N!$, is equal to $\exp\left[\sum_{n=1}^\infty a_nx^n/n!\right]$, see Eqs.~(\ref{apf2}) and~(\ref{apf3}), i.e. it is defined by the exponential generating function of the sequence $\{a_n\}$.

The above considerations mean that the zero-field square lattice Ising model is mathematically equivalent to a perfect gas of clusters. Of course, the alleged gas model referred to has nothing to do with the well-known lattice gas model which was studied by Yang and Lee \cite{1952LeeYang}, and in which the excluded volume effect must be taken into account. Moreover, even if one is skeptical as to whether one can ever determine the microscopic details of such a gas (i.e. details of its interparticle interactions), it can be shown that the mere idea of such a gas is very fruitful.

In order to show this, let us consider the energy distribution at a given temperature, i.e. the probability $P(N,x)$ of finding the system (both the Ising model and the perfect gas of energy-clusters model) in an equilibrium state with energy $2JN$ above the ground state. The energy distribution is simply given by:
\begin{equation}\label{pn1}
P(N,x)=\frac{2g(N)x^{N-1}}{Z(x)}.
\end{equation}
Substituting Eqs.~(\ref{apf3}) and (\ref{z1}) into this expression, and then using properties of Bell polynomials (see p.~135 in~\cite{bookComtet}), i.e.
\begin{eqnarray}\label{suppl1}
Y_N(\{cb^na_n\})=\sum_{k=1}^{N}c^kb^NB_{N,k}(\{a_n\}),
\end{eqnarray}
$P(N,x)$ can be written as (see Appendix \ref{sec4a}):
\begin{equation}\label{pn2}
P(N,x)=\frac{Y_N(\{a_nx^n\})/N!}{1+\sum_{N=1}^\infty Y_N(\{a_nx^n\})/N!}.
\end{equation}

Now, thinking in terms of a gas of independent energy-clusters and having in mind the general expression for the complete Bell polynomials, Eq.~(\ref{bell2}), the coefficients $\{a_nx^n\}$ after dividing them by $n!$ (to remove distinguishability of energy portions), may be interpreted as \emph{thermodynamic preferences} for clusters of size $n=1,2,\dots$. (To make this clear, the term 'thermodynamic preference' is used here for the product of the number of microscopic realizations of clusters, which consist of indistinguishable energy portions, $a_n/n!$, and the corresponding Boltzmann factor, $x^n$.) Then, using Eq.~(\ref{Anfig}), one can see that the introduced thermodynamic preferences strongly depend on temperature. For even values of $n$ one gets:
\begin{equation}\label{xxx}
\lim_{n\rightarrow\infty} \frac{a_n}{n!}x^n\simeq C\!\left(\frac{x}{x_c}\right)^n,
\end{equation}
from which it is easy to see that the passage through the critical point is related to the complete change in preferences on the size of energy clusters. Below the critical temperature, for $x<x_c$ (when the Ising model is in the ferromagnetic state), smaller clusters are characterized by higher preferences. In this temperature range, the preferences are an exponentially decreasing function of the cluster's size. On the other hand, above the critical temperature, for $x>x_c$ (when the Ising model is in the paramagnetic state), the preferences monotonically increase as a function of $n$. Phase transition occurs, when the preferences do not depend on clusters' size! This description in a vivid way illustrates the origins of phase transitions in the infinite systems. It also suggests, how finite-size systems modify this scenario by changing, above the critical point, a monotonically increasing sequence $\{a_nx^n/n!\}$ to unimodal $\{A_nx^n/n!\}$ .

Finally, Eq.~(\ref{xxx}) can be used to rewrite Eq.~(\ref{pn2}) in a compact way, i.e. for $x\leq x_c$ one has:
\begin{equation}\label{PEx}
P(N,x)\simeq\frac{\left(\frac{x}{x_c}\right)^{\!N}\!_{1}\!F_{1}(1\!-\!N;2;-C)} {C^{-1}\!+\!\sum_{N=1}^\infty\left(\frac{x}{x_c}\right)^{\!N}\!_{1}\!F_{1}(1\!-\!N;2;-C)},
\end{equation}
where $_{1}\!F_{1}(1-N;2;-C)$ is the so-called confluent hypergeometric function of the first kind \cite{wolfram1F1} (for details see Appendix  \ref{sec5a}), and the positive constant $C$, see Eq.~(\ref{Anfig}), can be determined from the condition of normalization of $P(N,x)$.

The last remark is related to the coefficients in the low-temperature series expansion of the partition function per spin, see~Eq.~(\ref{DombZ}),
\begin{equation}\label{OEIS}
0,\,0,\,0,\,1,\,0,\,2,\,0,\,5,\,0,\,14,\,0,\,44,\,0,\,152,\,0,\,566,\dots.
\end{equation}
It is clear that the coefficients can be easily obtained from Eqs.~(\ref{apf3}) and  (\ref{apf4}). In the Online Encyclopedia of Integer Sequences (OEIS) \cite{oeis} this sequence is catalogued under the number A002890. It is worth to mention that our approach not only presents exact formulae for the terms of this sequence but also provides fast method for calculating successive terms (see Appendix \ref{sec6a}).

\section{Summary}
In summary, in this paper we have used combinatorial formalism to obtain the exact low-temperature series expansion for the partition function of the two-dimensional zero-field $s=\frac{1}{2}$ Ising model on the infinite square lattice. We have shown that the phase transition in the Ising model can be explained through equivalence between the model and the perfect gas of energy clusters model, in which the passage through the critical point is related to the complete change in the thermodynamic preferences on the size of clusters. The combinatorial approach reported in this article is very general and can be easily applied to other models for which exact solutions are known.

\section{Acknowledgements}
The work has been supported from the National Science Centre in Poland (grant no. 2012/05/E/ST2/02300). GS also acknowledges the financial support from
internal funds of the Faculty of Physics at Warsaw University of Technology (grant no. 504/01425/1050/42.000100).

\appendix

\section{Low temperature series expansion of $-\beta f(\beta)$}\label{sec1a}

\par By substituting
\begin{equation}\label{s0}
x=e^{-2\beta J},
\end{equation}
and
\begin{equation}\label{s1}
p=p(\theta_1,\theta_2)=\cos\theta_1+\cos\theta_2,
\end{equation}
into Eq.~(\ref{f0}), the bulk free energy per site in the square lattice Ising model can be written as:\\
\begin{widetext}
\begin{eqnarray}\label{s2a}
-\beta f(\beta)\!=\!&\ln 2&\!+\;\frac{1}{8\pi^2} \int_{0}^{2\pi}d\theta_1\int_{0}^{2\pi}d\theta_2\ln\left[\left(\frac{x+x^{-1}}{2}\right)^2-\frac{-x+x^{-1}}{2}p  \right]\\\label{s2b}\!=\!&\ln 2&\!+\; \frac{1}{8\pi^2}\int_{0}^{2\pi}d\theta_1\int_{0}^{2\pi}d\theta_2 \ln\left[\frac{x^{-2}}{4}\left(x^4\!+\!2px^3\!+\!2x^2\!-\!2px\!+\!1\right)  \right]\\\label{s2c}\!=\!&\ln x^{-1}&\!+\;\frac{1}{8\pi^2} \int_{0}^{2\pi}d\theta_1\int_{0}^{2\pi}d\theta_2\ln\left(1\!-\!2px\!+\!2x^2\!+\!2px^3+x^4\right).
\end{eqnarray}
\end{widetext}

\par Next, the integrand function in Eq.~(\ref{s2c}) can be decomposed into a Taylor series as:\\
\begin{widetext}
\begin{eqnarray}\label{s3a}
\ln\left(1\!-\!2px\!+\!2x^2\!+\!2px^3+x^4\right)\!&\!\!=\!\!&\!\sum_{n=1}^\infty L_n(-2p,2\!\cdot\!2!,2p\!\cdot\!3!,4!)\frac{x^n}{n!} \\\label{s3b}\!&\!\!=\!\!&\!\sum_{n=1}^\infty \frac{x^n}{n!} \sum_{k=1}^n(\!-1)^{k-1}(k\!-\!1)!B_{n,k}(-2p,2\!\cdot\!2!,2p\!\cdot\!3!,4!),
\end{eqnarray}
\end{widetext}

where the so-called \emph{logarithmic polynomials} have been used, which are defined as (see Eq.~(5a), p.~140 in~\cite{bookComtet}):
\begin{eqnarray}\label{s4a}
\!&\!\!&\!\ln\left(\sum_{n=0}^\infty g_n\frac{x^n}{n!}\right)=\sum_{n=1}^\infty L_n\left(\{g_i\}\right)\frac{x^n}{n!}\\\label{s4b}\!&\!=\!&\!\sum_{n=1}^\infty \frac{x^n}{n!} \sum_{k=1}^n(-1)^{k-1}(k-1)!B_{n,k}\left(\{g_i\}\right),
\end{eqnarray}
where $B_{n,k}(\{g_i\})$ represent partial Bell polynomials, see Eq.~(\ref{bell2}).

Now, substituting Eq.~(\ref{s3b}) to (\ref{s2c}) one gets the general expression for the low temperature series expansion of the bulk free energy per site (cf.~Eq.~(\ref{f0})):
\begin{equation}\label{s5}
-\beta f(\beta)=-\ln x+\sum_{n=1}^\infty a_n\frac{x^n}{n!},
\end{equation}
where the expansion coefficients are given by:
\begin{align}\label{s6}
a_n=\frac{1}{8\pi^2}\sum_{k=1}^n(-1)^{k-1}(k-1)!\times\\ \times \int_{0}^{2\pi}d\theta_1 \int_0^{2\pi}d\theta_2 B_{n,k}(-2p,2\!\cdot\!2!,2p\!\cdot\!3!,4!).\nonumber
\end{align}

\par Eq.~(\ref{s6}) can be further simplified by using the explicit formula for partial Bell polynomials, Eq.~(\ref{bell2}), according to which the polynomial $B_{n,k}$ in Eq.~(\ref{s6}) can be written as:
\begin{align}
&B_{n,k}(-2p,2\!\cdot\!2!,2p\!\cdot\!3!,4!)\nonumber\\\label{s7a}&=n!\sum\limits_{d_1,d_2,d_3,d_4} \frac{(-2p)^{d_1}(2\!\cdot\!2)^{d_2}(2p\!\cdot\!3!)^{d_3}(4!)^{d_4}} {d_1!d_2!d_3!d_4! (1!)^{d_1}(2!)^{d_2}(3!)^{d_3}(4!)^{d_4}}\\\label{s7b} &=n!\sum\limits_{d_1,d_2,d_3,d_4} \frac{(-1)^{d_1}2^{d_1+d_2+d_3}}{d_1!d_2!d_3!d_4!}p^{d_1+d_3},
\end{align}
where the summation takes place over all integers $d_1,d_2,d_3,d_4\geq 0$, such that
\begin{equation}\label{s8a}
d_1\!+\!d_2\!+\!d_3\!+\!d_4=k,
\end{equation}
and
\begin{equation}\label{s8b}
d_1\!+\!2d_2\!+\!3d_3\!+\!4d_4=n.
\end{equation}
Now, after using Eqs.~(\ref{s7b}) and~(\ref{s8a}) in Eq.~(\ref{s6}) one gets the following expression for $a_n$:\\
\begin{widetext}
\begin{equation}\label{s9a}
a_n=-\frac{n!}{8\pi^2}\sum_{d_1,d_2,d_3,d_4} \frac{(-1)^{d_2+d_3+d_4}2^{d_1+d_2+d_3}} {(d_1+d_2+d_3+d_4)}\binom{d_1+d_2+d_3+d_4}{d_1,\,d_2,\,d_3,\,d_4}\int_{0}^{2\pi}\!\!d\theta_1 \int_0^{2\pi}\!\!d\theta_2\;p^{d_1+d_3},
\end{equation}
\end{widetext}
where the explicit summation over $k$ was omitted due to the fact that it is already included in the summation over the variables $d_1,d_3,d_3,d_4$ which now must only satisfy Eq.~(\ref{s8b}).

\par The last step towards the final expression for $a_n$ is to show that the double integral in Eq.~(\ref{s9a}) simplifies to:
\begin{equation}\label{s10}
\int_{0}^{2\pi}\!\!d\theta_1\int_{0}^{2\pi}\!\!d\theta_2\;p^l=
\begin{cases}0\quad & \mbox{for odd } l,\\4\pi^2\;2^{-l}\binom{l}{l/2}^2\quad & \mbox{for even } l,\end{cases}
\end{equation}
where $p$ is given by Eq.~(\ref{s1}). (For reasons of clarity, the detailed calculations leading to Eq.~(\ref{s10}) are not discussed here, but will be discussed in Sect.~\ref{sec2a} of this document.) Finally, by inserting Eq.~(\ref{s10}) into~(\ref{s9a}) one gets Eqs.~(\ref{An1}) and~(\ref{An2}) which are in use in the primary article: For odd values of $n$:
\begin{equation}\label{s11a}
a_n=0,
\end{equation}
and for even values ​​of $n$:
\begin{align}\label{s11b}
a_n=&\frac{n!}{2}\sum_{d_1,d_2,d_3,d_4}\binom{d_1+d_2+d_3+d_4}{d_1,\,d_2,\,d_3,\,d_4}\times\\&\times \frac{(-1)^{d_2+d_3+d_4-1}2^{d_2}}{d_1+d_2+d_3+d_4}\binom{d_1+d_3}{\frac{d_1+d_3}{2}}^2,\nonumber
\end{align}
where the summation takes place over all quadruple numbers $d_1,d_2,d_3,d_4\geq 0$, which satisfy conditions $d_1+2d_2+3d_3+4d_4=n$ and $d_1+d_3$ is even.

\section{Detailed calculations leading to Eq.~(\ref{s10})}\label{sec2a}

\par The double integral in Eq.~(\ref{s9a}) can be transformed as follows:
\begin{align}
&\int_{0}^{2\pi}\!\!d\theta_1\int_{0}^{2\pi}\!\!d\theta_2\;p^l\nonumber\\ \label{s12a}
&=\int_{0}^{2\pi}\!\!d\theta_1\int_{0}^{2\pi}\!\!d\theta_2\;p^l\\
&=\int_{0}^{2\pi}\!\!d\theta_1\int_{0}^{2\pi}\!\!d\theta_2\;(\cos\theta_1+\cos\theta_2)^l \\\label{s12b}&=2^l\int_{0}^{2\pi}\!\!d\theta_1\int_{0}^{2\pi}\!\!d\theta_2 \cos^l\left(\frac{\theta_1+\theta_2}{2}\right)\cos^l\left(\frac{\theta_1-\theta_2}{2}\right)
\\\label{s12c}&=2^{l-1}\int_{0}^{2\pi}\!\!du\;\cos^lu\int_{-2\pi}^{2\pi}\!\!dv\;\cos^lv \\\label{s12d}&=2^{l}\int_{0}^{2\pi}\!\!du\;\cos^lu \int_{0}^{2\pi}\!\!dv\;\cos^lv\\\label{s12e}&= 2^l\left(\int_0^{2\pi}\!\!d\theta\;\cos^l\theta\right)^2=2^l\phi_l^2,
\end{align}
where the integral $\phi_l$ satisfies the below expression
\begin{align}\label{s13a}
\phi_l&=\int_{0}^{2\pi}\!\!d\theta\;\cos^l\theta\\
&= \cos^{l-1}\theta\sin\theta\Big|_0^{2\pi}+ (l-1)\int_{0}^{2\pi}\!\!d\theta\;\cos^{l-2}\theta\;\sin^2\theta  \\\label{s13b}&=(l-1)\int_{0}^{2\pi}\!\!d\theta\;\cos^{l-2}\theta\left(1-\cos^2\theta\right) \\\label{s13c}&=(l-1)\phi_{l-2}-(l-1)\phi_{l}.
\end{align}
which leads to the following recursive equation:
\begin{equation}\label{s14}
\phi_l=\frac{l-1}{l}\;\phi_{l-2},\;\;\;\;\;\mbox{for}\;\;l=1,2,3\dots,
\end{equation}
with
\begin{equation}\label{s15}
\phi_0=\int_0^{2\pi}\!\!d\theta=2\pi\;\;\;\;\;\mbox{and}
\;\;\;\;\;\phi_1=\int_0^{2\pi}\!\!d\theta\;\cos\theta=0.
\end{equation}
Now, since the only solution of Eq.~(\ref{s14}) is
\begin{equation}\label{s16a}
\phi_l=0\;\;\;\;\;\mbox{for odd}\;\;\;\;\;l=1,3,5,\dots,
\end{equation}
and
\begin{equation}\label{s16b}
\phi_l=2\pi\frac{(l-1)!!}{l!!}\;\;\;\;\;\mbox{for even} \;\;\;\;\;l=0,2,4,\dots,
\end{equation}
Eq.~(\ref{s12e}) can be further simplified to:
\begin{eqnarray}\label{s17a}
\int_{0}^{2\pi}\!\!d\theta_1\int_{0}^{2\pi}\!\!d\theta_2\;p^l&=&
2^l(2\pi)^2\left(\frac{(l-1)!!}{l!!}\right)^2\\\label{s17b}
&=&4\pi^2\;2^l\left(\frac{l!}{l!!^2}\right)^2\\\label{s17c} &=&4\pi^2\;2^l\left(\frac{l!}{2^l(l/2)!^2}\right)^2\\\label{s17d} &=&4\pi^2\;2^{-l}{l\choose (l/2)}^2,
\end{eqnarray}
where the assumption that $l$ is even has been used. Eq.~(\ref{s17d}) exactly corresponds to Eq.~(\ref{s10}).

\begin{figure}[h]
\centerline{\epsfig{file=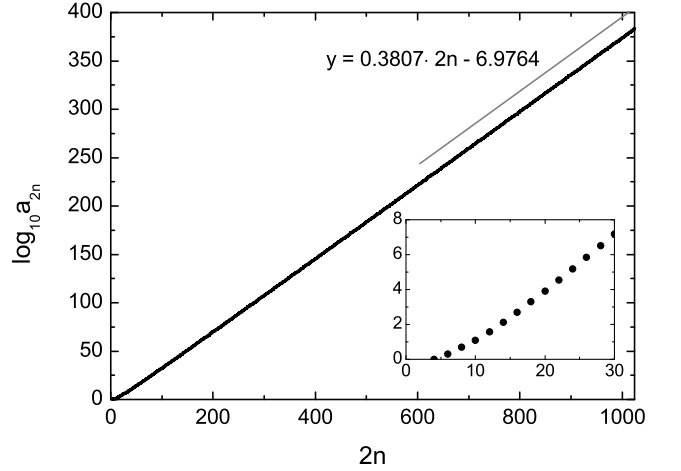,width=\columnwidth}}
\caption{Asymptotic behaviour of the sequence $a_{2n}/(2n)!$ vs $2n$.}
\label{figS2}
\end{figure}

\section{Asymptotic behaviour of the coefficients $a_{2n}/(2n)!$}\label{sec3a}
One can show that the coefficients in the low temperature series expansion of $-\beta f(\beta)$ (see Eqs.~(\ref{s5}),~(\ref{s11a}), and~(\ref{s11b})) have the asymptotic behaviour which is given by Eq.~(\ref{Anfig}):
\begin{equation}\label{s18}
\lim_{n\rightarrow\infty}\frac{a_{2n}}{(2n)!}=D\alpha^{2n},
\end{equation}
where $D$ is a positive constant, and (cf.~Eq.~(\ref{xxx}))
\begin{equation}\label{s19}
\alpha\simeq\frac{1}{x_c}=\exp\left[\frac{2J}{k_BT_c}\right]=\frac{1}{\sqrt{2}-1}.
\end{equation}
The log-linear plot of the coefficients $a_{2n}/(2n)!$ vs $2n$, which is shown in Fig.~\ref{figS2} illustrates this behaviour. The logarithm of $\alpha$:
\begin{equation}\label{s20}
\log_{10}\alpha\simeq 0.3828,
\end{equation}
corresponds to the slope of the line, $a=0.3807$,which is fitted to the results.

\section{Exact energy distribution $P(N,x)$ for the square lattice Ising model}\label{sec4a}

Eq.~(\ref{PEx}), which is exact in the limit of infinite lattice size, i.e. for $V\rightarrow\infty$, provides an excellent testbed for comparison of the exact infinite-volume results and the results of finite-size Monte Carlo methods (see e.g.\cite{1996PRLBeale, 2001PRLWang,bookLandau}).

\begin{figure}[h]
\centerline{\epsfig{file=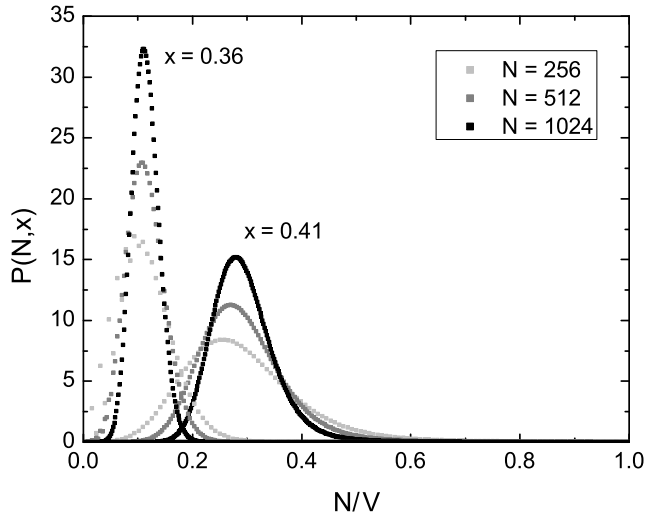,width=\columnwidth}}
\caption{Exact energy distribution $P(N,x)$ for the square lattice Ising model.}
\label{figS1}
\end{figure}

In Fig.~\ref{figS1}, the exact energy distribution $P(N,x)$, Eq.~(\ref{PEx}), is shown for three lattices of size: $V=256,\;512,\;1024$, and two different temperatures: $x=e^{-2\beta J}=0.36$ and $0.41$ (let us note that $x_c\simeq 0.414$).


\section{Detailed calculations leading to Eq.~(\ref{PEx})}\label{sec5a}

By using Eq.~(\ref{xxx}) and substituting $r$ for $\frac{x}{x_c}$, the numerator in Eq.~(\ref{pn2}) can be written as follows:
\begin{eqnarray}\label{s21a}
\frac{1}{N!}Y_N(\{a_nx^n\})&\simeq & \frac{1}{N!}Y_N(\{Cr^n n!\})\\\label{s21b}
&=&\frac{1}{N!}\sum_{k=1}^NB_{N,k}(\{Cr^n n!\})\\\label{s21c}
&=&\frac{1}{N!}\sum_{k=1}^NC^kr^N B_{N,k}(\{n!\}),
\end{eqnarray}
where the expression (\ref{suppl1}) has been used. Then, since the partial Bell polynomials with the coefficients: $1!,2!,3!\dots$ are equal to Lah numbers, 
\begin{equation}\label{s22}
B_{N,k}(1!,2!,3!,\dots)=\frac{N!}{k!}{N-1\choose k-1},
\end{equation}
Eq.~(\ref{s21c}) can be further simplified:\\
\begin{widetext}
\begin{eqnarray}\label{s23a}
\frac{1}{N!}Y_N(\{a_nx^n\})&\simeq &r^N\sum_{k=1}^N{N-1\choose k-1} \frac{C^k}{k!}\\\label{s23b} &\overset{N\gg 1}{\simeq}& r^NC\sum_{l=0}^{\infty}{N-1\choose l} \frac{C^l}{(l+1)!}\\\label{s23c} &=&r^NC\sum_{l=0}^{\infty}\left((-1)^l\frac{(N-1)!}{(N-1-l)!}\right) \left(\frac{1}{(l+1)!}\right)\frac{(-C)^l}{l!}\\\label{s23d}
&=&r^NC\sum_{l=0}^{\infty}\frac{(1-N)_k}{(2)_k}\frac{(-C)^l}{l!}
\\\label{s23e}&=&r^NC\;\;_{1}\!F_{1}(1\!-\!N;2;-C),
\end{eqnarray}
\end{widetext}

where $_{1}\!F_{1}(1-N;2;-C)$ is the so-called confluent hypergeometric function of the first kind [26], which is defined as:
\begin{eqnarray}\label{s24a}
_{1}\!F_{1}(a;b;z)&=&1+\frac{a}{b}z+\frac{a(a+1)}{b(b+1)}\frac{z^2}{2!}+\dots \\\label{s24b}&=&\sum_{k=0}^\infty\frac{(a)_k}{(b)_k}\frac{z^k}{k!},
\end{eqnarray}
where $(a)_k$ and $(b)_k$ are Pochhammer symbols.

Finally, by substituting Eq.~(\ref{s23e}) to (\ref{pn2}) one gets the energy distribution as given by Eq.~(\ref{PEx}).

\section{Mathematica routines}\label{sec6a}
In this section we present our Mathematica scripts which allow to calculate coefficients of the low-temperature expansion of the free energy, \{\texttt{an}\}, and coefficients of the expansion of the partition function, \{\texttt{Yn}\}.\\
Since the built-in Mathematica \texttt{BellY[]} function for calculating Bell polynomials works very slow, we implement Bell polynomials using the following recurrence formula (Eq. (3k) in [12])

\begin{equation*}
B_{n,k}(\{a_N\})=\sum_{l=k-1}^{n-1}\binom{n}{l}a_{n-l}B_{l,k-l}(\{a_N\}).
\end{equation*}

\begin{lstlisting}[language=Mathematica,caption={The coefficients of the free energy}]
In[1]:=   (*list of sets {d1,d2,d3,d4}, where d1+2*d2+3*d3+4*d4=n and d1+d3 is even *)
				  Belllist[n_] := Select[FrobeniusSolve[Range[4], n], EvenQ[#[[1]] + #[[3]]] &]
				  (*function of m={d1,d2,d3,d4} under the sum in Eq.(20) *)
				  ff[m_] := ((-1)^(m[[2]] + m[[3]] + m[[4]]) 2^m[[2]])/(m[[1]] + m[[2]] + m[[3]] + m[[4]])
					  		Multinomial[m[[1]], m[[2]], m[[3]], m[[4]]]
						  			Binomial[(m[[1]] + m[[3]]), (m[[1]] + m[[3]])/2]^2
				  (*final function for coefficients of free energy*)
				  a[n_] := - (n!/2)  If[OddQ[n],   0, Plus @@ (ff /@ Belllist[n])];
				  (*number of calculated coefficients*)
				  nN = 20;
				  (*list of the first nN coefficients divided by factorials*)
				  an = ParallelTable[a[n]/n!, {n, nN}]

Out[1]:= {0, 0, 0, 1, 0, 2, 0, 9/2, 0, 12, 0, 112/3, 0, 130, 0, 1961/4, 0, 5876/3, 0, 40871/5}
	
\end{lstlisting}

\begin{lstlisting}[language=Mathematica,caption={The coefficients of the partition function}]
In[2]:= (*the first nN coefficients NOT divided by factorials*)
					An = ParallelTable[A[n], {n, nN}];
					(*list of the coefficients of the partition function*)
					Yn = (Total /@ (Nest[MapThread[
					  Join[#1, {#2}] &, {#, 1/(Length@#[[1]] + 1)
						  Table[Sum[ Binomial[nn, l] An[[nn - l]]*
								#[[l, Length@#[[1]]]], {l,Length@#[[1]], nn - 1}], {nn, nN}]}] &, Partition[An, 1],
										    nN - 1])/Table[i!, {i, nN}])

Out[2]:= {0, 0, 0, 1, 0, 2, 0, 5, 0, 14, 0, 44, 0, 152, 0, 566, 0, 2234, 0, 9228}

\end{lstlisting}





\bibliography{DOS_Ising2dv2.uk}{}

\end{document}